\documentclass[prx, letterpaper, superscriptaddress, amsmath, amssymb, amssymb, reprint, floatfix]{revtex4-2}

\usepackage{graphicx}
\usepackage{dcolumn}
\usepackage{bm}
\usepackage{xcolor}
\usepackage{multirow}
\usepackage{upgreek}
\newcommand{\ket}[1]{\left\vert{#1}\right\rangle}
\newcommand{\bra}[1]{\left\langle{#1}\right\vert}
\newcommand{\braket}[1]{\left\langle{#1}\right\rangle}

\newcommand{\eqr}[1]{Eq.(\ref{#1})}
\newcommand{\tr}[1]{\mathbf{tr}\left\{#1\right\}}

\usepackage{textcomp}
\usepackage{xcite}
\usepackage{verbatim}
\usepackage[colorinlistoftodos]{todonotes}
\usepackage{microtype}

\usepackage{xr-hyper}
\usepackage{hyperref}
\makeatletter
\newcommand*{\addFileDependency}[1]{
  \typeout{(#1)}
  \@addtofilelist{#1}
  \IfFileExists{#1}{}{\typeout{No file #1.}}
}
\makeatother

\newcommand{\appropto}{\mathrel{\vcenter{
  \offinterlineskip\halign{\hfil$##$\cr
    \propto\cr\noalign{\kern2pt}\sim\cr\noalign{\kern-2pt}}}}}

\usepackage{booktabs}

\begin{document}


\preprint{APS/123-QED}

\title{Unified linear response theory of quantum electronic circuits}

\author{L. Peri}
\email{lp586@cam.ac.uk}
\affiliation{Quantum Motion, 9 Sterling Way, London N7 9HJ, United Kingdom}
\affiliation{Cavendish Laboratory, University of Cambridge, J.J. Thomson Avenue, Cambridge CB3 0HE, United Kingdom}
\author{M. Benito}
\email{monica.benito@uni-a.de}
\affiliation{Institute of Physics, University of Augsburg, 86159 Augsburg, Germany}

\author{C. J. B. Ford}
 \affiliation{Cavendish Laboratory, University of Cambridge, J.J. Thomson Avenue, Cambridge CB3 0HE, United Kingdom}
\author{M. F. Gonzalez-Zalba}
 \affiliation{Quantum Motion, 9 Sterling Way, London N7 9HJ, United Kingdom}

\date{\today}

\begin{abstract}
Modelling the electrical response of multi-level quantum systems at finite frequency has been typically performed in the context of two incomplete paradigms: (i) input-output theory, which is valid at any frequency but neglects dynamic losses, and (ii) semiclassical theory, which captures well dynamic dissipation effects but is only accurate at low frequencies. Here, we develop a unifying theory, valid for arbitrary frequencies, that captures both the quantum behaviour and the non-unitary effects introduced by relaxation and dephasing. The theory allows a multi-level system to be described by a universal small-signal equivalent circuit model, a resonant RLC circuit, whose topology only depends on the number of energy levels. 
We apply our model to a double quantum-dot charge qubit and a Majorana qubit, showing the capability to continuously describe the systems from adiabatic to resonant and from coherent to incoherent, suggesting new and realistic experiments for improved quantum state readout.
Our model will facilitate the design of hybrid quantum-classical circuits and the simulation of qubit control and quantum state readout.
\end{abstract}

\maketitle


\section{Introduction}

An accurate model of the electrical response of a quantum system is of paramount importance, particularly when engineering its manipulation and state readout within a larger classical circuit.
When an electrical signal perturbs the quantum dynamics of a mesoscopic quantum system, charge redistribution events will manifest as a (gate) current, which will propagate back into the classical world, where it will be measured \cite{Wallraff2004,Petersson2012,Samkharadzeeaar2018,Mi_2018,Derakhshan_2020,Burkard_Gullans_Mi_Petta_2020,vanDijk_2019}. This signal carries information of the dynamical properties of the system, which can be leveraged for characterization, tuning, and \textit{in-situ} readout of quantum information \cite{Sillanpaa2005,Oakesfast,Ciccarelli_Ferguson_2011,vigneau2022probing,vonhorstig2024electrical}.

Much has been studied about the electrical response of mesoscopic capacitors and resistors \cite{Buttiker_1993, Buttiker_1993_2, Nigg_Buttiker_2008,Gabelli_2006,Bruhat_Cottet_2016,Wang_Wang_Guo_2007}. However, a complete theoretical framework is still lacking in the quantum limit where the unitary dynamics of discrete levels and decoherence, caused by coupling to the environment, appear as dispersive and dissipative interactions to a connected classical circuit \cite{Ibberson_2021,Burkard_Petta_2016,Ibberson_2021,Persson_2010,Ciccarelli_Ferguson_2011,Peri2024beyondadiabatic}.
Thus far, modelling of such phenomena has been approached via two methods: 
(i) semiclassically, extending to quantum systems the treatment of classical high-frequency electronics \cite{Mizuta2017,Esterli_Otxoa_Gonzalez-Zalba_2019}, 
and (ii) with Input-Output theory, adapting quantum electro-dynamics (QED) methods to mesoscopic quantum devices operating in the radio-frequency (rf) or microwave range \cite{kohler_dispersive_2017,kohler_dispersive_2018,Benito_2017,Blais_Grimsmo_Girvin_Wallraff_2021}.
Both, however, only capture part of the picture, with the semiclassical models neglecting the complexity of the unitary dynamics beyond the adiabatic regime \cite{Shevchenko_2010,Ivakhnenko_2023}, and input-output theory lacking a complete description of dynamical dissipation processes.

In this work, we present a Lindblad perturbation formalism that bridges the gap between the two theories, and provides the missing pieces for a complete modelling of a quantum device at high frequency, including dynamic decoherence processes. 
Although the formalism will be of interest to many communities such as superconducting charge \cite{Srinivasan_2011,Yamamoto_2003}, semiconductor \cite{Samkharadzeeaar2018,Guo_2023,Burkard_2023,Bonsen_2023,Undseth_2023,Mi_2018} and Majorana qubits \cite{Smith_2020,Rainis_Loss_2012,Derakhshan_2020,Tsintzis_2024}, here we adopt the language typical of quantum dot (QD) systems \cite{Mizuta2017,vigneau2022probing,Esterli_Otxoa_Gonzalez-Zalba_2019}.  We will consider a quantum system in equilibrium excited by a (small) perturbation of the control voltages, and explore the link between its electrical response and susceptibility, arising from input-output formalism \cite{kohler_dispersive_2017,kohler_dispersive_2018}. The perturbation originates from a rf excitation of one of the gate voltages, $V_{\rm G}(t) = \delta V\cos{\omega t}$, and we work out its impact on the QD dynamics as well as the gate current it originates.

\begin{figure}[htb!]
  \centering
  \includegraphics[width = 0.9 \linewidth]{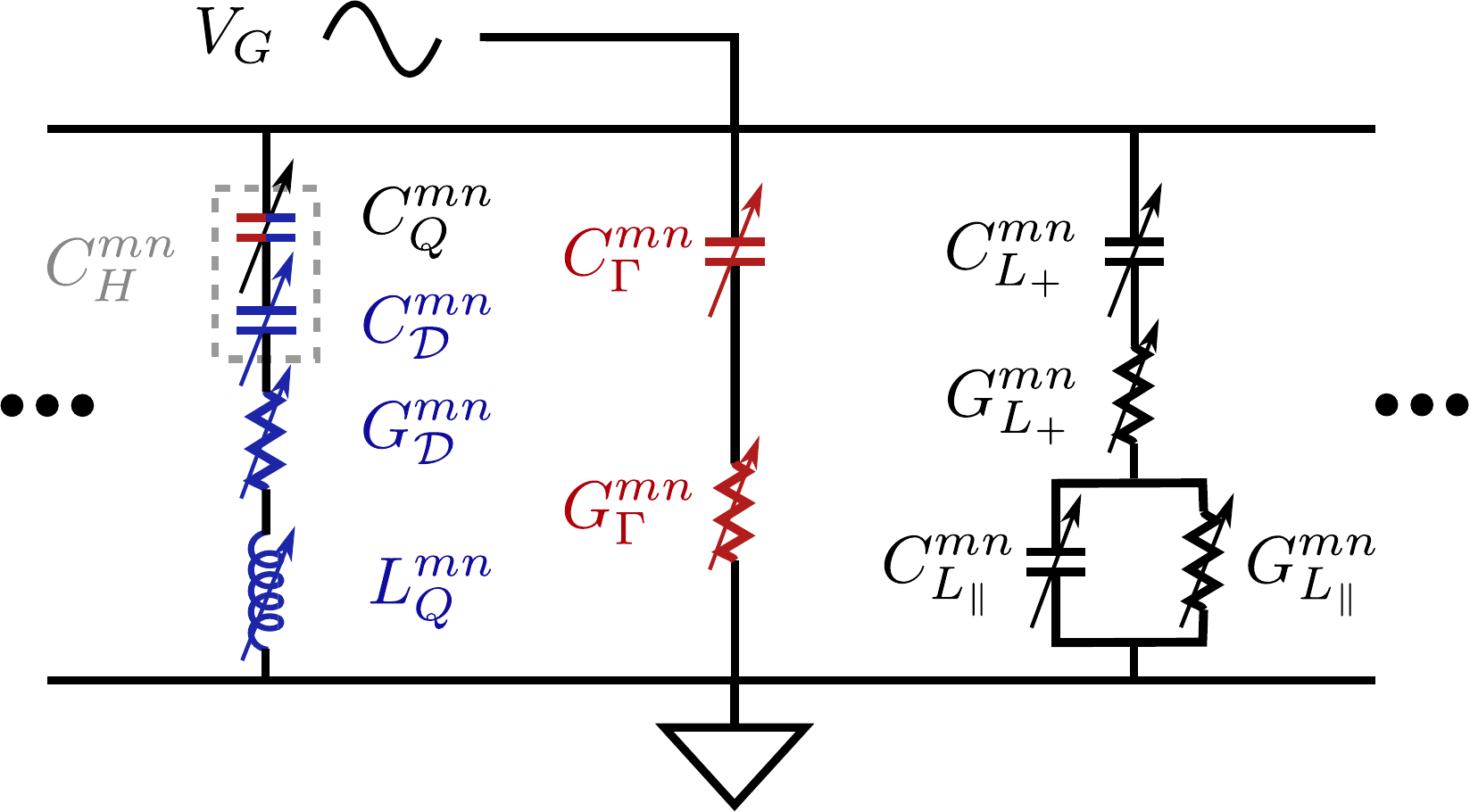}
  \caption{
    \textbf{Universal small-signal model of a quantum device embedded in a classical circuit.}
    Circuit elements are color-coded according to the description they originate from: input-output (blue) or semiclassical (red). Moreover, in this work we introduce a third branch (black), arising from a consistent quantum description of (detuning-dependent) dynamical decoherence.
    Notably, the input-output and semiclassical models overlap only in the description of the quantum capacitance (upper left).
    }
  \label{fig:topology}
\end{figure}

If the rf probe is taken to be small, the response of the system is \textit{linear}, and thus the gate current oscillates at the same frequency $\omega$ as the probe, allowing us to reframe the problem in terms of small-signal electrical analysis.
We find that, within the secular approximation, there exists a universal small-signal model of the system in terms of linear circuit elements, which consists of repeated fundamental units in \textit{parallel}, one for each \textit{pair} of levels (see Fig.~\ref{fig:topology}). Thus, the circuit topology only depends on the \textit{number} of levels described in the Hamiltonian, while the dynamical properties of the system only alter the \textit{values} of the components. 

In Fig.~\ref{fig:topology} we color-code the circuit elements according to the description they historically originate from: input-output (blue) or semiclassical (red). Moreover, a consistent description that includes both (detuning-dependent) decoherence and unitary evolution necessarily gives rise to a third set (black), which is introduced in this work and is dominant in the limit of very high decoherence. 
Notably, we show how the input-output and semiclassical models can be pictured to overlap in the description of the \textit{quantum capacitance}, which, as we shall see, is the adiabatic limit of the isolated quantum system. However, we shall also describe how the introduction of our novel term allows us to cast a new light on quantum capacitance, which can be seen additionally as the response of the system when all coherent processes are suppressed.

Our description of the response of the quantum circuit as linear circuit elements has the key benefit of enabling the inclusion of unitary \textit{quantum} behaviour in simple electronic circuit models. Moreover, we only make use of \textit{frequency-independent} voltage-dependent components, obviating the need for complex and slow Fourier-based nonlinear circuit analysis~\cite{Esterli_Otxoa_Gonzalez-Zalba_2019}.

This work is structured as follows. In Section~\ref{sec:Theory} we develop the formalism that leads to the universal small-signal model, starting from the simplicity of an isolated quantum system and later enhancing the model including decoherence in a Lindblad formalism.  
In Section \ref{sec:Examples} we apply our model to two example systems: a double quantum dot (DQD) charge qubit \cite{Mizuta2017,vigneau2022probing,Esterli_Otxoa_Gonzalez-Zalba_2019} and a QD coupled to topological Majorana modes \cite{Smith_2020,Rainis_Loss_2012,Derakhshan_2020,Tsintzis_2024}. We discuss our result in the context of the aforementioned paradigms present in the literature, demonstrating how our model continuously transitions between them in their respective regimes of validity. Moreover, we employ our formalism to suggest realistic experiments and potentially enhance current measurements.

\section{Universal small-signal model of a driven quantum system}
\label{sec:Theory}

In this Section we describe the theoretical basis of our formalism, relating the electrical response of a perturbed quantum system to its (charge) susceptibility, whose link leads us to a universal small-signal model of a driven quantum system.

The first necessary step is to describe the driven system quantum-mechanically, and thus define the gate current in terms of a quantum observable. For simplicity and ease of notation, we consider the case of a single gate driving the system (see App.~\ref{app:MultiGate}), but allow for an arbitrary number of energy levels (i.e., of orbital, valley or spin origin). The perturbation of the control voltages is related to the energy detuning of the system via a dimensionless parameter $\alpha$ (the gate lever arm). Thus, in terms of energy detuning, $\varepsilon(t) = \varepsilon_0 + \delta \varepsilon \cos{\omega t}$, where the detuning oscillation amplitude reads $\delta \varepsilon = \alpha e \delta V$ \cite{vigneau2022probing}.

To the quantum system, (perturbed) gate voltages appear as a time-dependent Hamiltonian of the form
\begin{equation}
 H = H_0 + \delta \varepsilon \Pi \cos{\omega t}
 \label{eq:H_t}
\end{equation} 
\noindent 
where $H_0$ is the unperturbed Hamiltonian, and $\Pi$ is the \textit{dipole} operator that couples the quantum system to the gate, which reads
\begin{equation}
  \Pi = \frac{d}{d \varepsilon} H_0 = \frac{1}{\alpha e}\frac{d}{d V_{\rm G}} H_0,
  \label{eq:Pi_def_main}
\end{equation}
\noindent
which can be understood physically by noting that, in our definition, $|\braket{\psi|\Pi|\psi}|^2$ determines how much each level $\ket{\psi}$ is capacitively coupled to the gate. 
Throughout this work, we shall assume that $\Pi$ is \textit{constant} with respect to gate voltages (i.e., $\frac{d^2}{d \varepsilon^2} H_0 = 0$) as happens for most physical systems, such as QD arrays in the constant-interaction approximation \cite{vanderWiel_2002}.

Using the previous physical intuition, we can write the gate current as
\begin{equation}
  I_{\rm G} = \alpha e \frac{d}{d t} \braket{\Pi}(t) ,
  \label{eq:I_G_q_main}
\end{equation}
\noindent
where $\alpha e \braket{\Pi}(t)$ can be interpreted as the (time-dependent) polarization charge at the gate. 
Notably, we stress that the operator $\Pi$ describes both the perturbation of the QD detuning and collects the back-action on the gate, as one would expect since they originate from the \textit{same} capacitive coupling.

\subsection*{Admittance and Quantum Capacitance}

Since the response of the system is \textit{linear}, one can consider the small-signal admittance of the QD system,
\begin{equation}
  Y = \frac{\int_0^{\frac{2\pi}{\omega}} e^{i \omega t} I_G(t) dt}{\int_0^{\frac{2\pi}{\omega}} e^{i \omega t} V_G(t) dt} = \frac{\omega \alpha e}{\pi \delta \varepsilon}\int_0^{\frac{2\pi}{\omega}} e^{i \omega t} I_G(t) dt,
  \label{eq:Y_gen}
\end{equation}
\noindent
which simplifies if we write explicitly
\begin{equation}
  \braket{\Pi} = \sum_m p_m \braket{\phi_m\left|\frac{d}{d \varepsilon} H_0\right|\phi_m} = \sum_m p_m \frac{d}{d \varepsilon} E_m,
  \label{eq:pi_dec_prov}
\end{equation}
\noindent
where $\ket{\phi_m}$ are the eigenstates of the system (given by $H_0\ket{\phi_m} = E_m \ket{\phi_m}$) and $p_m$ are their probabilities (see App.~\ref{P_eq}), while the second step follows from the Hellman-Feynman theorem.

If the rf perturbation is slow enough to be considered adiabatic, and neglecting for now thermal redistribution of probability, we can write 
\begin{equation}
    I_{\rm G} (t)  = \alpha e \sum_m p_m \frac{d}{d t}\left( \frac{d}{d \varepsilon} E_m (t) \right),
\end{equation}
\noindent from which we obtain via Taylor expansion
\begin{equation}
 Y=i \omega C_{\rm Q} = -i \omega(\alpha e)^2 \sum_m p_m \frac{d^2}{d \varepsilon^2} E_{m}\bigg|_{\varepsilon=\varepsilon_0},
  \label{eq:C_q}
\end{equation}
\noindent where $C_{\rm Q}$ is commonly known as the quantum capacitance \cite{Mizuta2017,Esterli_Otxoa_Gonzalez-Zalba_2019,park_adiabatic_2020}.
If the perturbation also mixes the eigenstates, however, we will have other contributions to the admittance, which, most generally, must be written as a complex admittance $Y(\omega)$ \cite{Mizuta2017,Esterli_Otxoa_Gonzalez-Zalba_2019,vigneau2022probing}.
The real part represents \textit{resistive} effects, and thus energy dissipation caused by the interaction of the QD with the environment, while the imaginary part, other than the quantum capacitance, originates from the redistribution of probability, either thermally or because of diabatic transitions, hence the name \textit{tunnelling} capacitance in the literature \cite{Mizuta2017}. In App.~\ref{C_param_der} we sketch out a similar argument to extend this form of calculation of gate current and admittance to the general case.

\subsection*{Isolated Quantum System}

We now consider the case of the isolated quantum system, the evolution of which is perfectly unitary. When unperturbed, its evolution is described by the Von Neumann equation as
\begin{equation}
  \frac{d}{dt} \rho = -i \left[H_0, \rho\right] ,
  \label{eq:VonNeumann}
\end{equation}
\noindent
where we have introduced the density matrix $\rho$ to describe the system, and take $\hbar =1$. If the system is in the steady state with $\rho \equiv \rho_{\rm ss}$, i.e., $\left[H_0, \rho_{\rm ss}\right] = 0$, we obtain the gate current caused by the perturbation via the Kubo formula \cite{Albert_Bradlyn_Fraas_Jiang_2016}
\begin{equation}
  I_{\rm G}(t) = \frac{\alpha e \delta \varepsilon}{i} \frac{d}{dt} \int^{t}_{-\infty} d \tau \cos{(\omega \tau)}
  \tr{\left[\Pi(t, \tau), \Pi\right]\rho_{\rm ss}} ,
  \label{eq:Ig_kubo}
\end{equation}
\noindent
where $\Pi(t, \tau)$ represents the operator in interaction picture. In \eqr{eq:Ig_kubo}, however, we can recognize the \textit{susceptibility} of the system \cite{kohler_dispersive_2017,kohler_dispersive_2018}
\begin{equation}
  \chi(t, \tau) = \frac{1}{i}\tr{\left[\Pi(t, \tau), \Pi\right]\rho_{\rm ss}} \Theta(t-\tau) ,
  \label{eq:chi_def}
\end{equation}
\noindent
from which we can immediately write the gate current 
\begin{equation}
  I_{\rm G}(t) = \alpha e \delta \varepsilon \frac{d}{dt} \int^{+\infty}_{-\infty} d \tau \cos{(\omega \tau)} \chi(t, \tau),
  \label{eq:Ig_chi}
\end{equation}
\noindent
linking the susceptibility of the system with its electrical response. Lastly, we note that in the small-signal regime the dynamics is not a function of $t$ and $\tau$ separately, but only of the time difference $t - \tau$, as is the susceptibility. Therefore, we can define the susceptibility of the system in reciprocal space $\chi(\omega)$ by taking its Fourier transform \cite{kohler_dispersive_2018}. We can now use the causality requirement of \eqr{eq:chi_def}, and thus the well-known result that 
$\chi(-\omega) = \chi(\omega)^*$, to write the electrical response as 
\begin{align}
  &I_{\rm G}(t) = \alpha e \omega \delta \varepsilon \left(\Re\left[\chi(\omega)\right] \sin{\omega t} + \Im\left[\chi(\omega)\right] \cos{\omega t}\right)
  \label{eq:I_from_chi}\\
  &Y(\omega) = 2i \omega (\alpha e)^2 \chi^*(\omega).
  \label{eq:Y_from_chi}
\end{align}

This property follows from the fact that the operator determining the gate current is \textit{the same} as the operator that determines the perturbation. Thus \eqr{eq:Ig_chi} is a consequence of the interaction between the classical circuit and the quantum system. For a unitary evolution (App.~\ref{Chi_calc}), the susceptibility takes the well-known form \cite{kohler_dispersive_2018}
\begin{equation}
  \chi(\omega) = \sum_{m, n} \left(p_m-p_n\right)\frac{\big|\braket{\phi_m\big|\Pi\big|\phi_n}\big|^2}{\omega - (E_n - E_m)}.
  \label{eq:chi_H}
\end{equation}
Firstly, $\chi(\omega)$ is purely real, and thus the system's response is purely reactive, as required by \eqr{eq:VonNeumann}. To compare this result with \eqr{eq:C_q}, it is interesting to take the adiabatic limit of \eqr{eq:chi_H}. Recalling our assumption that $\frac{d^2}{d\varepsilon^2} H_0 = 0$ and making use again of Hellman-Feynman \cite{park_adiabatic_2020}, we find
\begin{equation}
  \frac{d^2}{d \varepsilon^2}  E_m = \frac{1}{2} \sum_{n} \frac{\big|\braket{\phi_m\big|\Pi\big|\phi_n}\big|^2}{E_m - E_n},
  \label{eq:CQ_as_sum}
\end{equation}
\noindent
which leads to the expected result 
\begin{equation}
  2 (\alpha e)^2 \lim_{\omega \rightarrow 0} \chi^*(\omega) =  C_{\rm Q}.
\end{equation}
\noindent

Equation~(\ref{eq:chi_H}), however, shows how, even for an isolated quantum system, the finite-frequency response cannot simply be modelled by the quantum capacitance $C_{\rm Q}$. Unlike in a simple capacitor, the susceptibility diverges exactly at resonance, as one would expect from general theory \cite{kohler_dispersive_2018}. This behaviour can be better understood if we notice that every pair of levels is summed over \textit{twice} in \eqr{eq:chi_H}. 
The hermiticity of $\Pi$ allows us, thanks to Foster's theorem \cite{Foster_1924,Nigg_2012,Solgun_2014}, to synthesize the admittance of the system in the much more insightful form of an $LC$ resonator, with
\begin{equation}
  \begin{aligned}
  &Y(\omega) = \sum_{E_m>E_n} \left(\frac{1}{i \omega C_{\rm Q}^{mn}} + i \omega L_{\rm Q}^{mn}\right)^{-1} ,  \\
  & C_{\rm Q}^{mn} =  \frac{(\alpha e)^2 \eta_{mn}}{\left(E_m-E_n\right) ^ 2}
  \hspace{0.05\textwidth} L_{\rm Q}^{mn} =  \frac{1}{(\alpha e)^2 \eta_{mn}} ,
  \end{aligned}
  \label{eq:Y_LC_isol}
\end{equation}
\noindent
where we have defined
\begin{equation}
  \eta_{mn} = 2 \left(p_n-p_m\right) \left(E_m-E_n\right)\big|\braket{\phi_m\big|\Pi\big|\phi_n}\big|^2.
\end{equation}
Firstly, we note how this confirms our previous adiabatic derivation of the quantum capacitance, as, thanks to the summation rule in \eqr{eq:CQ_as_sum}, we find
\begin{equation}
  \sum_{E_m>E_n} C_{\rm Q}^{mn} = C_{\rm Q}.
\end{equation}
At finite frequencies, however, the full diabatic quantum response implies the existence of a \textit{quantum inductance} $L_{\rm Q}^{mn}$, as observed in the literature for a single QD \cite{Frey_2012,Wang_Wang_Guo_2007}. The system is thus best modelled as a set of perfect LC resonators \textit{in parallel}, one per \textit{pair} of levels and resonant at the energy splitting.

\subsection*{Open Quantum System}

It is of interest to go beyond the isolated system and consider the presence of decoherence because of the coupling between the QDs and the environment. The quantum dynamics of an open system can be described via the Lindblad Master Equation (LME) \cite{manzano_short_2020}
\begin{align}
  & \frac{d}{dt} \rho = \mathcal{L}_0 \rho = -i \left[H_0, \rho\right] + \sum_l \Gamma_l \mathcal{D}\left(L_l\right)\rho \label{eq:LME} \\
  & \mathcal{D}\left(L_l\right) \rho = L_l\rho L_l^\dagger - \frac{1}{2}\left\{L_l^\dagger L_l , \rho \right\},
\end{align}
\noindent
where $\Gamma_l$ are called tunnel rates, while $\mathcal{D}\left(L_l\right)$ is the superoperator causing decoherence effects, described by the \textit{jump} operators $L_l$. In this case, \eqr{eq:Ig_chi} remains valid provided that we now redefine the susceptibility as \cite{Albert_Bradlyn_Fraas_Jiang_2016}
\begin{equation}
  \chi(t, \tau) = \tr{\Pi e^{\mathcal{L}_0(t-\tau)}\delta \mathcal{L}\rho_{\rm ss}} \Theta(t-\tau) ,
  \label{eq:chi_def_L}
\end{equation}
\noindent
where $e^{\mathcal{L}_0(t-\tau)}$ is the unperturbed propagator of the (non-unitary) dynamics, and $\delta \mathcal{L}$ is the (small) perturbation to the Liouvillian $\mathcal{L}_0$ caused by the excitation. Given the linearity of the LME, we can divide the perturbation as $ \delta \mathcal{L} =  \delta \mathcal{L}_{\rm H} +  \delta \mathcal{L}_{\Gamma} + \delta \mathcal{L}_{\rm L}$, where we consider the effect of the rf excitation on the Hamiltonian, the tunnel rates, and the jump operators respectively. Their time evolution can be expressed to first order in the perturbation as 
\begin{equation}
  \begin{aligned}
  &\Gamma_l(t) = \Gamma_l^{0} + \delta \varepsilon \frac{d \Gamma_l}{d \varepsilon} \cos{\omega t} + \mathcal{O}(\delta \varepsilon^2)\\
  &L_l(t) = L_l^{0} + \delta \varepsilon \frac{d L_l}{d \varepsilon} \cos{\omega t}+ \mathcal{O}(\delta \varepsilon^2) .
  \end{aligned}
  \label{eq:perturbation}
\end{equation}

Linearity in \eqr{eq:chi_def_L} lets us write the total susceptibility as 
$
  \chi(\omega) = \chi_{\rm H}(\omega) + \chi_\Gamma(\omega) + \chi_{\rm L}(\omega)
$
and thus, from \eqr{eq:Y_from_chi}, the admittance can be synthesized as \cite{Zinn_1952,Montgomery_Dicke_Purcell_1987,Nigg_2012}
\begin{equation}
  Y = 2i \omega (\alpha e)^2 \chi^*(\omega) = Y_{\rm H} + Y_\Gamma + Y_{\rm L}.
\end{equation}

Notably, this can be thought of in electrical terms as a \textit{parallel} combination of the three contributions (Fig.~\ref{fig:topology}), which we now discuss in detail.
The result in \eqr{eq:chi_def_L} is valid for any Liouvillian, not necessarily in Lindblad form. However, for the sake of simplicity, we assume that the dynamics of the system is well described by a phenomenological model including relaxation ($T_1$) and pure dephasing ($T_\phi$) processes. Thus, we can write 
\begin{equation}
  \begin{aligned}
    \mathcal{L}_0 \rho &= -i \left[H, \rho\right] + \sum_{E_m>E_n} \Gamma_\phi^{mn} \mathcal{D}\left(\tau^z_{nm}\right) \rho \\
     & + \sum_{E_m>E_n} \Gamma_+^{mn} \mathcal{D}\left(\tau^x_{mn}\right) \rho
  + \Gamma_-^{mn} \mathcal{D}\left(\tau^x_{nm}\right) \rho,
\end{aligned}
\label{eq:L_def_gen}
\end{equation}
\noindent
where we have defined 
\begin{align}
  &\tau^x_{mn} = \ket{\phi_m} \bra{\phi_n}, & \tau^z_{nm} = \frac{\ket{\phi_m}\bra{\phi_m} - \ket{\phi_n}\bra{\phi_n}}{\sqrt{2}}.
\end{align}

\subsubsection*{Hamiltonian Admittance}
The simplest term is the contribution due to the direct perturbation of the Hamiltonian. Similarly to \eqr{eq:chi_def}, this reads
\begin{equation}
  \delta \mathcal{L}_{\rm H} \rho = -i \delta \varepsilon \cos{\omega t} \left[\Pi, \rho\right].
\end{equation}
In App.~\ref{Chi_calc} we show that considering the Lindbladian in \eqr{eq:L_def_gen} leads to the simple modification of \eqr{eq:chi_H} \cite{kohler_dispersive_2018}
\begin{equation}
  \chi_{\rm H}(\omega) = \sum_{m, n} \left(p_m-p_n\right)\frac{\big|\braket{\phi_m\big|\Pi\big|\phi_n}\big|^2}{\omega - (E_n - E_m) + i \Gamma^{mn}_{T_2}},
  \label{eq:chi_L}
\end{equation}
\noindent
where 
\begin{equation}
  \Gamma_{T_2}^{mn} = \Gamma_{\phi}^{mn} + \frac{\Gamma_{+}^{mn} + \Gamma_{-}}{2}
\end{equation}
is the rate of decay of off-diagonal terms in the density matrix \cite{kohler_driven_2005}.
Therefore, the effect of metastable states is to give $\chi_{\rm H}(\omega)$ a small imaginary part with respect to the isolated system, allowing the open quantum system to dissipate energy, thus curing the divergence of the response exactly at resonance.
Notably, we find a diabatic \textit{decoherence-induced} resistance, discussed in the literature of mesoscopic capacitors \cite{Wang_Wang_Guo_2007,Buttiker_1993,Buttiker_1993_2,Cottet_Mora_Kontos_2011,Bruhat_Cottet_2016,Gabelli_2006,Nigg_Buttiker_2008} but notably absent in the literature of equivalent-circuit models of QD systems \cite{Mizuta2017,Esterli_Otxoa_Gonzalez-Zalba_2019,vigneau2022probing}. 

Exploring this further, the Choi-Kraus theorem guarantees that $\Gamma^{mn}_{T_2} = \Gamma^{nm}_{T_2}$~\cite{manzano_short_2020}. Thus, we find the admittance related to $\chi_{\rm H}$ as
\begin{equation}
  Y_{\rm H}(\omega) =  \sum_{E_m>E_n} \left(\frac{1}{G_\mathcal{D}^{mn}} + \frac{1}{i \omega C_{\rm H}^{mn}} + i \omega L_{\rm Q}^{mn}\right)^{-1} ,
  \label{eq:Y_H} 
\end{equation}
\noindent where we have introduced
\begin{equation}
  G_{\mathcal{D}}^{mn} = \frac{(\alpha e)^2 \eta_{mn}}{2\Gamma^{mn}_{T_2}} 
  \hspace{0.075\textwidth} C_{\mathcal{D}}^{mn} = \frac{ (\alpha e)^2 \eta_{mn}}{\left(\Gamma^{mn}_{T_2}\right)^{2}}
\end{equation}
\noindent
and $ C_{\rm H}^{mn} =  \left(\frac{1}{C_{\rm Q}^{mn}} + \frac{1}{C_{\mathcal{D}}^{mn}}\right)^{-1}$.
As for the isolated system (\eqr{eq:Y_LC_isol}), the diabatic response of the Hamiltonian component can be pictured as arising from resonators in parallel, one for each pair of levels. These are now, however, RLC resonators, as the presence of decoherence introduces both a \textit{resistive} component $G_{\mathcal{D}}^{mn}$ and an \textit{additional capacitance} $C_{\mathcal{D}}^{mn}$ in series with the quantum terms (Fig.~\ref{fig:topology}). Notably, the frequency pulling of $C_{\mathcal{D}}^{mn}$ cancels out the damping exactly, and the resonance frequency remains unaltered at the value $\hbar \omega_{\textnormal{res}} = E_m-E_n$.

\subsubsection*{Sisyphus Admittance}
To first order in the excitation, the perturbation of the Lindbladian due to the tunnel rates takes the form
\begin{equation}
  \delta \mathcal{L}_\Gamma \rho = \delta \varepsilon \cos{\omega t} \sum_l \frac{d \Gamma_l}{d \varepsilon}\mathcal{D}\left(L_l^0\right) \rho.
\end{equation}
In the literature, this effect has been termed \textit{Sisyphus} processes, responsible for the resistance of the same name, and has been identified as the main cause of dynamical dissipation in QD systems \cite{Mizuta2017,Esterli_Otxoa_Gonzalez-Zalba_2019,Ciccarelli_Ferguson_2011,Persson_2010,vigneau2022probing}.
The physical cause of Sisyphus processes stems from the dependence of the tunnel rates on energy \cite{vigneau2022probing}.
The simplest way to describe time-dependent excitation is the instantaneous eigenvalue approximation (IEA), where the functional form of the time-independent master equation is calculated for the instantaneous eigenvalues of $H(t)$ \cite{Yamaguchi_Yuge_Ogawa_2017}. Thus $\Gamma_\pm^{mn} (t) = \Gamma_\pm \left(E_m(t) - E_n(t)\right)$.
For a small perturbation, this causes the population to relax to equilibrium (App.~\ref{P_eq}) with rate 
\begin{equation}
  \Gamma_{T_1}^{mn} = \Gamma_+(E_m-E_n) + \Gamma_-(E_m-E_n).
\end{equation}
However, along one excitation cycle, the resulting modulation of the rates can give rise to \textit{excess} relaxation, 
\begin{equation}
  \frac{d \Gamma_{\pm}^{mn}}{d \varepsilon} = \left(\braket{\phi_n\big|\Pi\big|\phi_n} - \braket{\phi_m\big|\Pi\big|\phi_m}\right) \frac{d}{dE} \Gamma_\pm (E_m - E_n) ,
\end{equation}
which is manifested as a resistive term \cite{Ciccarelli_Ferguson_2011,Persson_2010,Esterli_Otxoa_Gonzalez-Zalba_2019}. 

In App.~\ref{Sisyphus}, we show how the electrical response arising from the Sisyphus process can be written as
\begin{align}
  &Y_\Gamma(\omega) = \sum_{E_m>E_n} \left( \frac{1}{G_\Gamma^{mn}} + \frac{1}{i \omega C_\Gamma^{mn}}\right) ^ {-1} ,
  \label{eq:Y_sis} 
\end{align}
\noindent where we have defined 
\begin{align}
  & G_\Gamma^{mn} = (\alpha e)^2 \sigma_{mn} ,
  \hspace{0.05\textwidth} C_\Gamma^{mn} = \frac{ (\alpha e)^2 \sigma_{mn}}{\Gamma_{T_1}^{mn}}\\
  & \sigma_{mn} = \left(\frac{d \Gamma^{mn}_+}{d \varepsilon} p_m - \frac{d \Gamma^{mn}_-}{d \varepsilon} p_n\right) \left(\braket{\phi_n\big|\Pi\big|\phi_n} - \braket{\phi_m\big|\Pi\big|\phi_m}\right). \nonumber
\end{align}
This is a generalization of the Sisyphus conductance derived for the DQD or the single electron box \cite{Mizuta2017,Esterli_Otxoa_Gonzalez-Zalba_2019,vigneau2022probing}. Notably, \eqr{eq:Y_sis} shows how Sisyphus processes generally include also a \textit{reactive} component, thus inviting us to refer to a Sisyphus \textit{admittance} that includes both the Sisyphus conductance and the tunnelling capacitance \cite{Esterli_Otxoa_Gonzalez-Zalba_2019}, as they are manifestations of the same physical process. Electrically, the series $RC$ combinations behave as high-pass filters, one for each pair of levels, with corner frequency $\Gamma_{T_1}^{mn}$ (Fig.~\ref{fig:topology}), with the clear physical significance of \textit{suppressing} unitary processes that would happen more slowly than the lifetime of the levels.

\subsubsection*{Hermes Admittance}
The last term contributing to the susceptibility is the perturbation of the jump operators. In the notation of \eqr{eq:perturbation}, this reads
\begin{equation}
  \delta \mathcal{L}_{\rm L} \rho = \delta \varepsilon \cos{\omega t} \sum_l \Gamma_l^0 \mathcal{D}'(L_l) \rho ,
\end{equation}
\noindent where we have defined
\begin{equation}
  \mathcal{D}'(L_l) \rho =  L_l^0 \rho \frac{d L_l^\dagger}{d \varepsilon} - \frac{1}{2}\left\{\frac{d L_l^\dagger}{d \varepsilon} L_l^0 , \rho \right\} + h.c.
\end{equation}
Within the IEA, jump operators are usually defined in the instantaneous eigenbasis (IEB) in which $H(t)$ is instantaneously diagonal \cite{oakes2022quantum,Yamaguchi_Yuge_Ogawa_2017}. Thus, $ L_l(t) = W^\dagger(t) L_l^{\rm IEB}W(t) $, where the rotation matrix $W(t)$ is such that $ W^\dagger(t) E(t) W(t)= H(t) $, with $E(t)$ the diagonal matrix containing the instantaneous energies. 
Incidentally, this term is a \textit{requirement} to get a valid Lindbladian \cite{Yamaguchi_Yuge_Ogawa_2017,Albash_Boixo_Lidar_Zanardi_2012,Ikeda_Chinzei_Sato_2021,Mori_2023}. This is clear as, if the tunnel rates depend on the energies of the instantaneous eigenstates, the jump operators must describe transitions between such \textit{same} levels. Thus, a consistent quantum treatment of Sisyphus processes \textit{must} account for $\delta \mathcal{L}_{\rm L}$ as well.
In App.~\ref{Hermes}, we show how the response reads
\begin{equation}
  \chi_{\rm L}(\omega) = i\sum_{m, n} \Gamma^{mn}_{T_2} \frac{p_m-p_n}{E_m - E_n} \frac{\big|\braket{\phi_m\big|\Pi\big|\phi_n}\big|^2}{\omega - (E_n - E_m) + i \Gamma^{mn}_{T_2}}. 
  \label{eq:chi_HE}
\end{equation}
\noindent
This term is a novel effect introduced in this work, and includes the effect of both relaxation ($T_1$) and pure dephasing ($T_\phi$) processes, which combine to give the total 
decoherence rate as $T_2^{-1} = T_\phi^{-1} + \left(2T_1\right)^{-1}$.

Firstly, we note the stark similarity between $\chi_{\rm H}$ and $\chi_{\rm L}$, which is to be expected since the perturbation of the jump operators stems from the changes of the Hamiltonian and its eigendecomposition.
Similarities and differences become evident if we analyze the response from the circuit point of view. After some algebra, we can write
\begin{align}
    &Y_{\rm L}(\omega) =\sum_{E_m>E_n}  \left(\frac{1}{G_{\rm L_+}^{mn}} + \frac{1}{i \omega C_{\rm L_+}^{mn}} +  \frac{1}{G_{\rm L_\parallel}^{mn} + i \omega C_{\rm L_\parallel}^{mn}}\right)^{-1} ,
    \nonumber \\[1ex]
  \label{eq:Y_HE} 
\end{align}
\noindent
where we define
\begin{align}
  &C_{\rm L_+}^{mn} = \gamma C_{\rm H}^{mn} 
  \hspace{0.1\textwidth} G_{\rm L_+}^{mn} = 2 \gamma G_{\mathcal{D}}^{mn} \\
  &C_{\rm L_\parallel}^{mn}  = - \gamma C_{Q}^{mn}
  \hspace{0.085\textwidth}G_{\rm L_\parallel}^{mn}  = -2 \gamma^2 G_{\mathcal{D}}^{mn} \nonumber
\end{align}
\noindent
and $\gamma = \left(\frac{\Gamma_{T_2}^{mn}}{E_m-E_n}\right)^2$.
Notably, also for this term, the admittance corresponds to circuits in parallel, one for each pair of levels (Fig.~\ref{fig:topology}), and the values of the equivalent circuit components are proportional (via $\gamma$) to their Hamiltonian counterparts.
However, we note that the inductance in $\chi_{\rm H}$ is here replaced by a parallel combination of a (negative) resistor and a (negative) capacitor. 
Thus, similarly to the Sisyphus term, the $\chi_{\rm L}$ does \textit{not} resonate \cite{Alavi_Mahdi_Payne_Howey_2017}, but, rather, acts to \textit{dampen} resonances caused by the Hamiltonian term.
We notice how all the terms in $\chi_{\rm L}$ depend on $\gamma$, i.e., on the (squared) ratio between the coherent \textit{quantum beat} between levels and the decay of their coherent superposition. 
As a matter of fact, we shall see in the subsequent section how this term dominates over the \textit{quantum} (Hamiltonian) contribution when the system decoheres faster than its natural frequency ($\gamma>1$), and in this regime it is responsible for the recovery of the semiclassical limit.
Expanding on this, it is in fact possible to show (after some algebra) that 
\begin{equation}
  \lim_{\gamma\rightarrow \infty } \left(Y_{\rm H} + Y_{\rm L}\right) = \sum_{E_m>E_n} i \omega C_{\rm Q}^{mn} ,
  \label{eq:Y_semicl_limit}
\end{equation}
\noindent 
which casts a new light on the concept of quantum capacitance. This can now be seen as \textit{either} the adiabatic limit of the isolated system (as argued above) \textit{or} as the response of the quantum system in the limit of infinitely fast decoherence (which the literature sometimes defines as the semiclassical limit \cite{kohler_floquet-markovian_1997,Rudner_Lindner_2020}). 
Physically, these two views are reconciled by noticing how in both regimes it is impossible for the quantum system to escape its steady state, either because of the adiabaticity of the drive or the suppression of coherent superposition and thus unitary processes.

Lastly, we address the issue of naming the new term $\chi_{\rm L}$. As we have seen, this depends on the velocity of decoherence compared to the intrinsic beat of the Hamiltonian dynamics. Thus, in keeping with the mythological theme established by Sisyphus, it seems only fitting to name $\chi_{\rm L}$ the \textit{Hermes} susceptibility, from the Greek deity of velocity and mischievousness.

\section{Circuit model of example systems}
\label{sec:Examples}

In this Section, we employ the formalism derived thus far and we showcase its capabilities on two example systems: (i) a DQD charge qubit and (ii) a Majorana qubit formed by a QD coupled to two topological Majorana modes.

\subsection*{Charge Qubit}

Firstly, we discuss a charge qubit in a DQD, chosen because it is the simplest coupled two-level system and the (semiclassical) Sisyphus admittance and adiabatic quantum capacitance are well-known from the literature \cite{Mizuta2017,Esterli_Otxoa_Gonzalez-Zalba_2019}.
In particular, we use the Lindblad equivalent of the semiclassical model employed in Refs.~\cite{Mizuta2017} and \cite{Esterli_Otxoa_Gonzalez-Zalba_2019}, where the Hamiltonian in the charge basis of the two QDs reads 
\begin{equation}
  H(t) = \frac{1}{2}(\varepsilon_0 + \delta \varepsilon \cos{\omega t}) \sigma_z + \frac{\Delta}{2} \sigma_x = H_0 + \delta \varepsilon \Pi \cos{\omega t}.
\end{equation}
\noindent
Here, $\Delta/2$ is the tunnel coupling.  Relaxation is phenomenologically introduced as \cite{oakes2022quantum}
\begin{equation}
  \begin{aligned}
    &L^{\rm IEB}_+ = \ket{e}\bra{g} \hspace{0.07\textwidth} \Gamma_+(\Delta E) = \Gamma_0 n(\Delta E)\\
    &L^{\rm IEB}_- = \ket{g}\bra{e} \hspace{0.07\textwidth} \Gamma_-(\Delta E) = \Gamma_0 \left[n(\Delta E) + 1\right] ,
  \end{aligned}
  \label{eq:DQD_jump}
\end{equation}
\noindent
where $\Delta E = \sqrt{\Delta^2 + \varepsilon^2}$ is the energy difference between the ground ($\ket{g}$) and excited ($\ket{e}$) states, and $n(\Delta E) = \left(e^{\Delta E/k_{\rm B} T} -1\right)^{-1}$. 
As is common in the literature \cite{Esterli_Otxoa_Gonzalez-Zalba_2019,Derakhshan_2020}, we take $\Gamma_0$ to be independent of energy. The small-signal admittance of the DQD can be found by direct application of the above equations, taking 
\begin{equation}
  \Gamma_{T_1} = \Gamma_+(\Delta E) + \Gamma_-(\Delta E) = \Gamma_0 \left(2n(\Delta E) +1\right)
\end{equation}
\noindent and, in the absence for now of a dephasing term, $\Gamma_{T_2} = \frac{1}{2}\Gamma_{T_1}$.
In particular, considering Eqs.~(\ref{eq:Y_H}), (\ref{eq:Y_sis}), and (\ref{eq:chi_HE}), one obtains
\begin{align}
    & Y_{\rm H} = i \frac{\alpha e^2}{2}\frac{\Delta^2}{\Delta E} \frac{\Gamma_0}{\Gamma_{T_1}} \frac{\omega}{\Delta E^2 + \left(\Gamma_{T_2} + i \omega\right)^2} \label{eq:Y_H_DQD}\\
    & Y_\Gamma = \frac{\alpha e^2}{4 k_{\rm B} T} \frac{\varepsilon_0^2}{\Delta E^2} \frac{\Gamma_0^2}{\Gamma_{T_1}} \frac{\omega}{\omega - i \Gamma_{T_1}} \sinh^{-2}\left(\frac{\Delta E}{2 k_{\rm B} T}\right) \label{eq:Y_sis_DQD}\\
    & Y_{\rm L} = i\frac{\alpha e^2}{2}\frac{\Gamma_{T_2} \Delta^2}{\Delta E^3} \frac{\Gamma_0}{\Gamma_{T_1}} \frac{\omega(\Gamma_{T_2} + i \omega)}{\Delta E^2 + \left(\Gamma_{T_2} + i \omega\right)^2} , \label{eq:Y_He_DQD}
\end{align}
\noindent
where, notably, $\braket{e|\Pi|g} = \Delta/\Delta E$ is the dipole matrix element, and the polarization in the energy basis reads $p_g - p_e = \Gamma_0 / \Gamma_{T_1}$.

\begin{figure}[htb]
  \centering
  \includegraphics[width = 0.95\linewidth ]{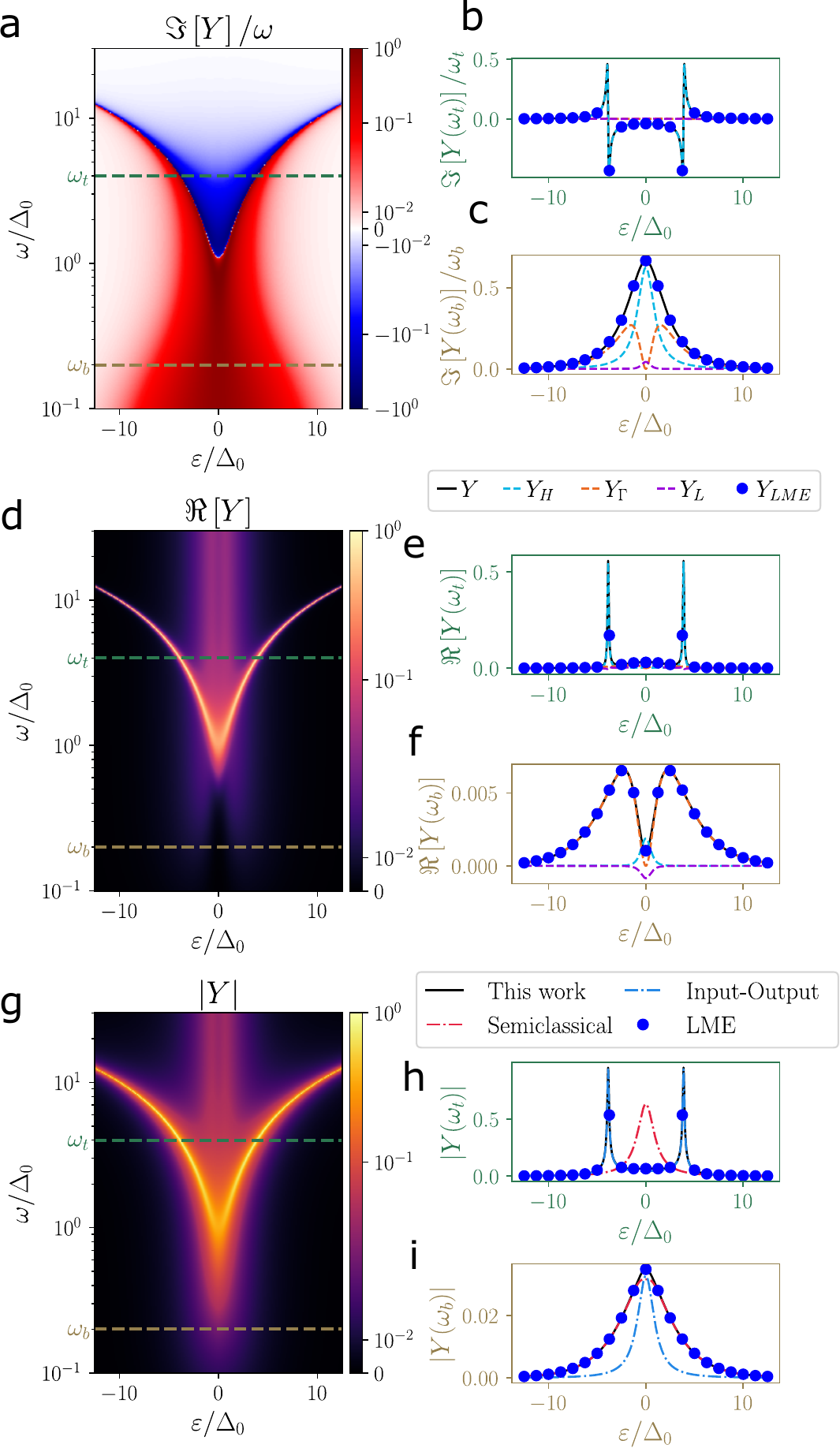}
  \caption{\textbf{Coherent charge qubit.} 
  Capacitance (a-c), conductance(d-f) and absolute value of the admittance (g-i) of the DQD charge qubit for increasing frequency at $k_{\rm B} T = 2.5 \Delta$ and $\Gamma_0 = \Delta_0/10$. The line cuts are taken in the adiabatic ($\omega_b = 0.2 \Delta$) and resonant ($\omega_t = 4 \Delta$) regimes.
  Panels b-c and e-f show the contribution to real and imaginary parts of the admittance from the Hamiltonian (cyan), Sisyphus (orange) and Hermes (pink) components, while panels h-i show a comparison between our model and the semiclassical model \cite{Esterli_Otxoa_Gonzalez-Zalba_2019} and input-output theory \cite{kohler_dispersive_2018}. We additionally carry out a brute-force integration of the LME (blue dots), showing perfect agreement with our analytical model. Color bars are linear for the $\pm 5$\% of the values on either side of zero and logarithmic outside.}
  \label{fig:fig2_omega}
\end{figure}

In Fig.~\ref{fig:fig2_omega}, we showcase our model by considering the effects of increasing the excitation frequency, comparing them to the semiclassical \cite{Esterli_Otxoa_Gonzalez-Zalba_2019} and input-output \cite{kohler_dispersive_2018} results. We choose the case of high temperature ($k_{\rm B} T > \Delta$) to enhance the contribution of the Sisyphus term.
For completeness, we additionally carry out a brute-force integration of the LME (blue dots), showing complete agreement with our analytical model.
To begin with, we note how, as expected, the semiclassical admittance is recovered for $\omega\rightarrow 0$. In particular, we see how the width of the peak becomes broader than the mere quantum capacitance (Fig.~\ref{fig:fig2_omega}c,i) due to the Sisyphus term, in agreement with the semiclassical theory \cite{Esterli_Otxoa_Gonzalez-Zalba_2019,vigneau2022probing} and experimental observations \cite{DiCarlo_2004,Hu_2007}.
Notably, in this work we obtain the same functional form for Sisyphus resistance and tunnelling (Sisyphus) capacitance (\eqr{eq:Y_sis_DQD}) as Ref.~\cite{Esterli_Otxoa_Gonzalez-Zalba_2019}, as expected since this term arises from the perturbation of scalars. More specifically, it is possible to show (App.~\ref{Sisyphus}) that if $\rho_{\rm ss}$ is diagonal (as it typically is in thermal equilibrium \cite{Albert_Bradlyn_Fraas_Jiang_2016,kohler_floquet-markovian_1997,kohler_dispersive_2018}), then $\delta \mathcal{L}_{\Gamma} \rho_{\rm ss}$ is \textit{also} diagonal. 
Therefore, this term of the LME is perfectly equivalent to a semiclassical (scalar) master equation. In passing, we note that this is \textit{not} the case for $\delta \mathcal{L}_{\rm H}$ or $\delta \mathcal{L}_{\rm L}$.
In contrast to the semiclassical theory, input-output theory ignores the dynamical relaxation process, thus predicting at low excitation frequencies a narrower, temperature-independent \cite{kohler_dispersive_2018} peak, which cannot faithfully reproduce the LME (Fig.~\ref{fig:fig2_omega}i). 
This discrepancy is further highlighted when one considers the resistive component of the admittance (Fig.~\ref{fig:fig2_omega}f), where the Sisyphus term dominates and we obtain the characteristic two lobes \cite{Esterli_Otxoa_Gonzalez-Zalba_2019} caused by the dependence of (thermal) relaxation on $\Delta E$. The semiclassical model, however, fails to correctly predict the additional conductance peak at zero detuning, where the Sisyphus term vanishes and dissipation is dominated by $G_{\mathcal{D}}$ and by the Hermes term.

The situation is reversed when $\omega > \Delta$. (Fig.~\ref{fig:fig2_omega}c,f,h). We observe the Rabi-induced peak splitting caused by the inductive component in $\chi_{\rm H}$, while within the two Rabi wings ($\omega > \Delta E$) the reactive component changes sign. Both phenomena are predicted by Input-Output theory \cite{kohler_dispersive_2018,Benito_2017} and circuit QED \cite{Manucharyan_Baksic_Ciuti_2017,Toida_Nakajima_Komiyama_2013}, and have a simple physical interpretation. 
The splitting of the peak is caused by the strong response (divergent in the non-dissipative limit) of the quantum system when driven exactly at resonance. Within the wings, the system is driven faster than its natural response frequency, and thus the charge \textit{lags} behind the excitation, giving rise to a negative reactance. 
Both observations have been confirmed by experimentally \cite{Ibberson_2021,Frey_2012,Zhou_2022}, while they are obviously absent from the semiclassical (adiabatic) model.
A similar peak splitting due to vacuum Rabi is observed also in the conductance of the charge qubit, the response dominated by $Y_{\rm H}$ and sharply peaked at resonance. 
Similarly to the capacitive response at resonance, the large conductance can be understood by the system being efficiently driven by the excitation. 
This constant interplay of excitation of the quantum system and relaxation through the bath causes a net energy dissipation, which manifests electrically as a resistance. 
This is further clarified thinking of the process in the formalism of circuit QED, where instead of charge dynamics, we are invited to keep track of photons exchanged with the cavity, which in turn drive the quantum system \cite{kohler_dispersive_2018}. If the drive is resonant with the energy splitting, photons can efficiently be absorbed by the system being excited in a coherent superposition, which then decays back to equilibrium, leaking energy into the environment.

This demonstrates how our formalism is able to continuously morph between the two regimes, always remaining in agreement with the LME, showcasing how this work \textit{unifies} the semiclassical and input-output descriptions.

\begin{figure}[htb]
  \centering
  \includegraphics[width = 0.99 \linewidth ]{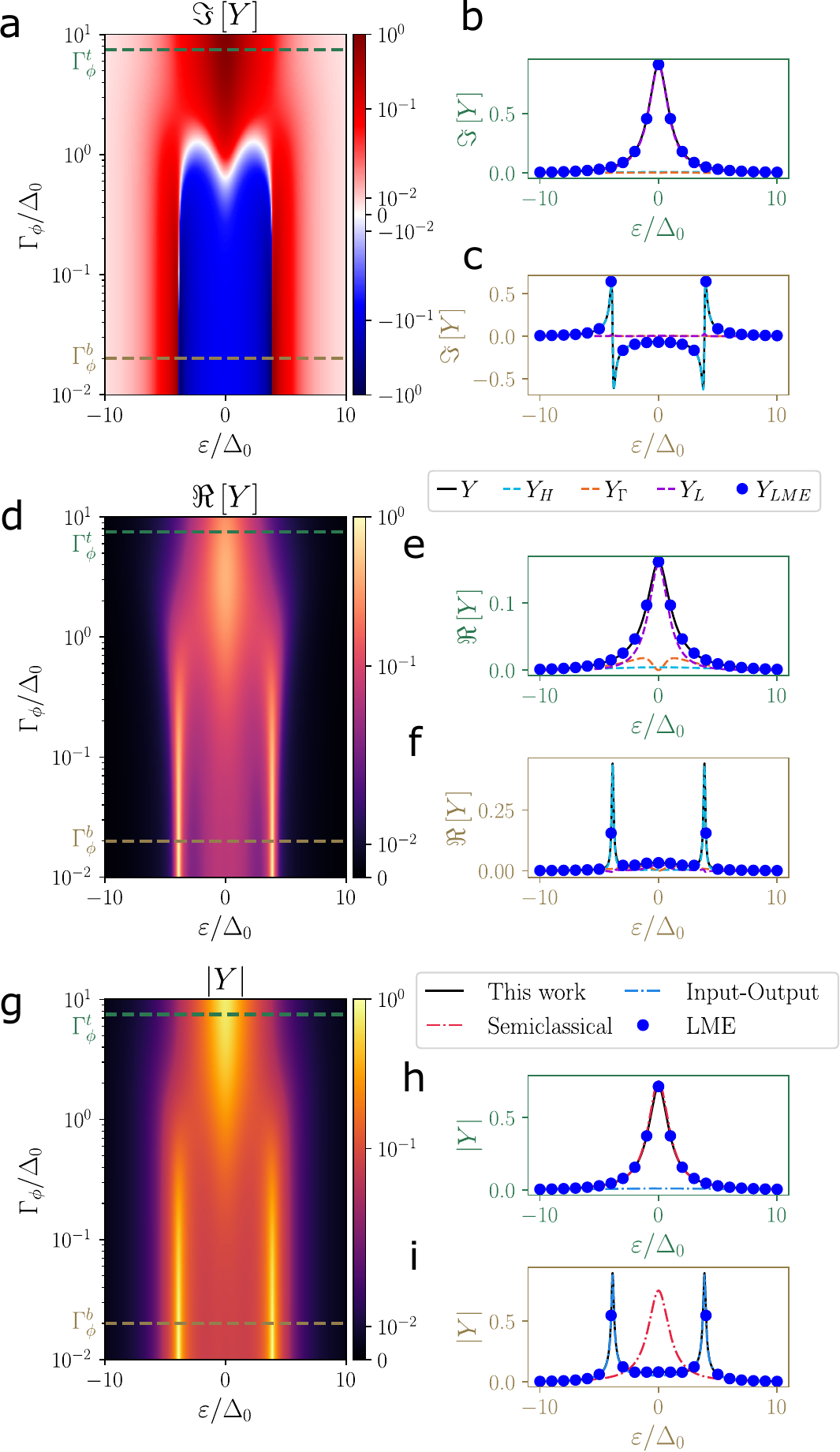}
  \caption{\textbf{Effect of dephasing.} Capacitance (a-c), conductance(d-f) and absolute value of the admittance (g-i) of the DQD charge qubit in the resonant regime ($\omega = 4 \Delta$) for increasing dephasing rate $\Gamma_\phi$ at $k_{\rm B} T = 2.5 \Delta$, and $\Gamma_0 = \Delta_0/10$. The line cuts are taken in the low-dephasing ($\Gamma_\phi = 0.02 \Delta$) and high-dephasing ($\Gamma_\phi = 7.5 \Delta$) regimes. Panels b-c and e-f show the contribution to real and imaginary part of the admittance from the Hamiltoinan (cyan), Sisyphus (orange) and Hermes (pink) components, while panels h-i show a comparison between our model and the semiclassical model \cite{Esterli_Otxoa_Gonzalez-Zalba_2019} and input-output theory \cite{kohler_dispersive_2018}. We additionally carry out a brute-force integration of the LME (blue dots), showing perfect agreement with our analytical model. Scale bars are linear within $\pm 5$\% and logarithmic outside.}
  \label{fig:fig3_dephasing}
\end{figure}

Lastly, we comment on the new Hermes term, which becomes important when, unlike in Fig.~\ref{fig:fig2_omega}, $\Gamma_{T_2}$ is comparable with $\Delta E$.
To showcase this effect, we introduce a pure dephasing term of the form
\begin{equation}
  L^{\rm IEB}_z = \frac{\ket{g}\bra{g}-\ket{e}\bra{e}}{\sqrt{2}} \hspace{0.1\textwidth} \Gamma_z(E) = \Gamma_\phi ,
  \label{eq:DQD_phi}
\end{equation}
\noindent
which has no semiclassical equivalent. For simplicity, we take $\Gamma_\phi$ also independent of energy. Thus, the only necessary modification of Eqs.~(\ref{eq:Y_H_DQD}-\ref{eq:Y_He_DQD}) is redefining $\Gamma_{T_2} = \Gamma_\phi + \Gamma_{T_1}/2$. 
Consequently, varying $\Gamma_\phi$ decouples the effect of $\chi_{\rm H}$ and $\chi_{\rm L}$ from the Sisyphus term. 

In Fig.~\ref{fig:fig3_dephasing}, we consider the DQD in the resonant regime (Fig.~\ref{fig:fig2_omega}c,f,i) and show the evolution of the admittance for increasing $\Gamma_\phi$.
For low decoherence rates, we see how the addition of a dephasing term does not qualitatively alter the admittance, which still shows a strong response sharply peaked at resonance (Fig.~\ref{fig:fig3_dephasing}c,f,i) and a change in the sign of the reactance (Fig.~\ref{fig:fig3_dephasing}f), a phenomenon well described by input-output theory (Fig.~\ref{fig:fig3_dephasing}i). 
When $\Gamma_{T_2}$ approaches $\Delta E$, however, not only do the resonant peaks become broader, but the Hermes contribution starts to become more relevant, in the shape of a single peak. 

This trend continues up to the point where the system decoheres faster than the timescales of the unitary dynamics ($\Gamma_{T_2}>\Delta E$).
In this case, we see how the Hermes term begins to dominate over the Hamiltonian (Fig.~\ref{fig:fig3_dephasing}b,e), and the Rabi peaks disappear completely, the admittance having the shape now of a more conventional single peak centered at $\varepsilon = 0$. In fact, when $\Gamma_{T_2} \gg \Delta E$, we completely recover the semiclassical prediction, as one would expect from the limit of very high decoherence (Fig.~\ref{fig:fig3_dephasing}i).
Unlike in the adiabatic regime, however, we see that the resistive component \textit{also} takes the shape of a zero-centred peak rather than the Sisyphus lobes (Fig.~\ref{fig:fig2_omega}i). This is the physical manifestation of the fact that in this regime the dominant process is not the dynamical relaxation, but, rather, the dynamical loss of coherence (and thus \textit{leakage} of quantum information), which is most efficient at $\varepsilon = 0$, where the dipole of the system is largest.

Therefore, we stress how the inclusion of the Hermes term is not only crucial to correctly reproduce the results from the LME, but, most importantly, to correctly recover the semiclassical limit (\eqr{eq:Y_semicl_limit}), as desirable in any complete and consistent modelling effort.

\subsection*{Majorana Qubit}

As a second example, we discuss the equivalent admittance of a Majorana qubit \cite{Flensberg_2011,Gharavi_Hoving_Baugh_2016,Karzig_2017,Kitaev_2001,Knapp_Karzig_Lutchyn_Nayak_2018,Plugge_Rasmussen_Egger_Flensberg_2017,Smith_2020,Rainis_Loss_2012,Derakhshan_2020,Tsintzis_2024}. Typically, a Majorana qubit is based on a topological one-dimensional system housing two Majorana zero modes, which make up the computational basis. 
For concreteness, in this work we concentrate on the simplest incarnation of such a qubit, in which the topological modes are coupled to an ancillary QD \cite{Flensberg_2011,Derakhshan_2020}. However, our formalism can be generalized naturally to more recent Majorana qubit proposals \cite{Plugge_Rasmussen_Egger_Flensberg_2017,Smith_2020}, as well as similar non-abelian systems that lack formal topological protection, such as Kitaev chains and ``poor man's Majorana'' bound states \cite{Tsintzis_2024}.

\begin{figure}[htb!]
  \centering
  \includegraphics[width = 0.99 \linewidth ]{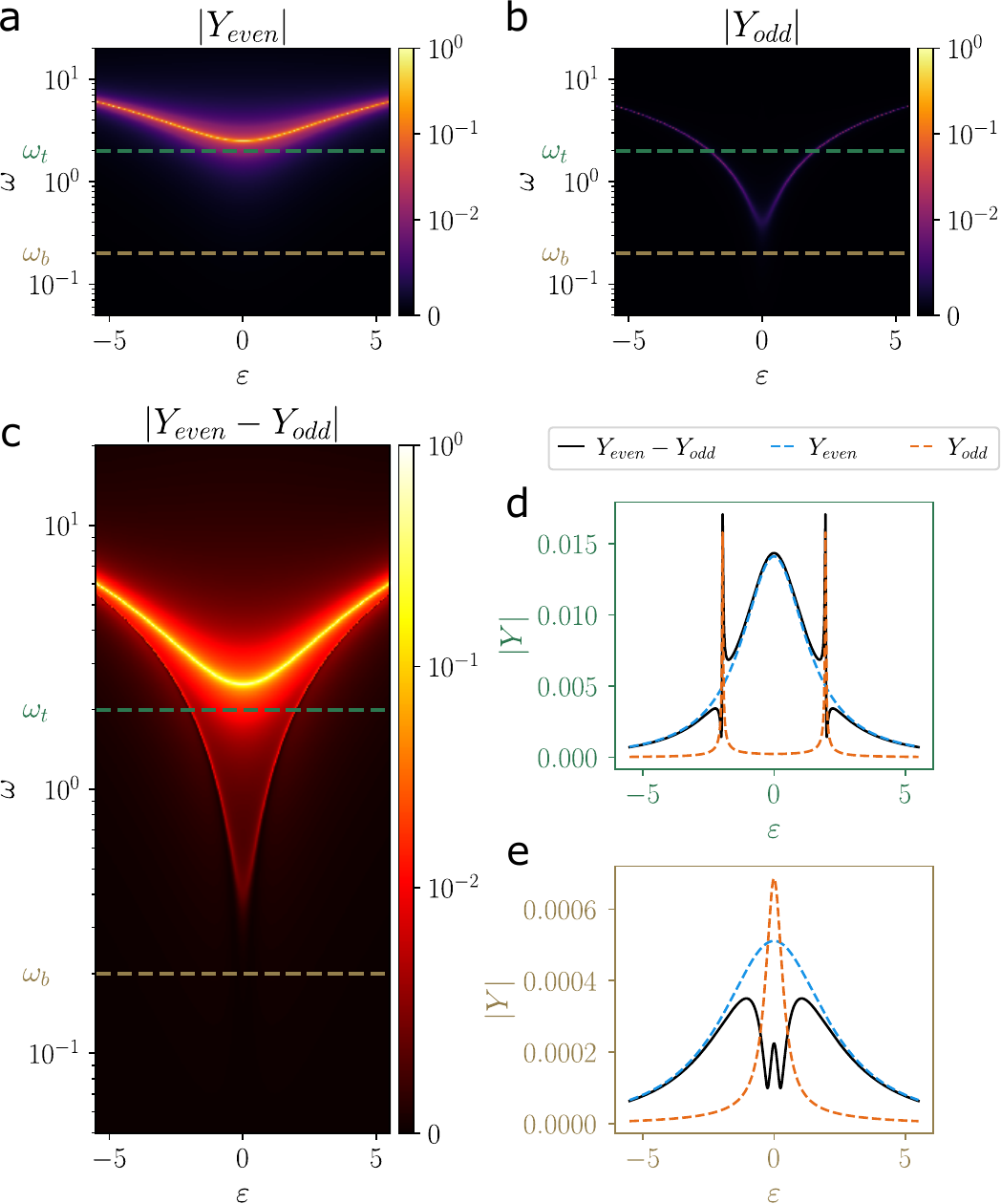}
  \caption{\textbf{Majorana qubit.} (a-b) Absolute value of the admittance of the auxiliary QD in case of the even (a) and odd (b) joint Majorana parity. (c) Absolute value of the admittance difference between the two parity states, which directly translates to the visibility of the readout of the two states. (d-e) $|Y_{\textnormal{even}} - Y_{\textnormal{odd}}|$ in the resonant (d) and adiabatic (e) regime, showing the benefit of performing readout at $\Delta_{\textnormal{odd}} < \omega < \Delta_{\textnormal{even}}$ to suppress the admittance of the odd state at $\varepsilon = 0$, which instead has a strong response at $\varepsilon_{\textnormal{odd}} = \sqrt{\omega - \Delta_{\textnormal{odd}}}$ where it is resonant with the excitation.
  Color bars are linear below $10^{-2}$ and logarithmic above.}
  \label{fig:fig4_Majorana}
\end{figure}

When an auxiliary QD is coupled to two Majorana zero modes, it can be used to read out the joint Majorana parity, i.e., to sense the occupation of the (non-local) fermion composed by the two Majorana modes \cite{Derakhshan_2020,Tsintzis_2024}. 
The effective low-energy Hamiltonian of such a system simply reads 
\begin{equation}
  H_p = \frac{\varepsilon}{2} \sigma_z + \frac{\Delta_p}{2}\sigma_x ,
\end{equation} 
\noindent
where $\varepsilon$ is the detuning of the ancillary QD, and $p \in \{\textnormal{even},\textnormal{odd}\}$ is the joint Majorana parity \cite{Derakhshan_2020}. $\Delta_p$ represents the Majorana-QD coupling in each parity state, and can be tuned experimentally by changing barrier voltages and the magnetic flux threading the device. Most importantly, generally $\Delta_{\textnormal{even}} \neq \Delta_{\textnormal{odd}}$, and, thus, the parity state can be measured through the admittance of the coupled QD \cite{Derakhshan_2020}.
For the sake of simplicity, in the following discussion we will only include decoherence processes (modelled as in Eqs.~(\ref{eq:DQD_jump})) within the even and odd sectors, while ignoring processes that can alter the Majorana parity, making the (realistic) assumption that they are negligible within the typical timescales of qubit readout \cite{Karzig_Cole_Pikulin_2021,Karzig_2017}.

We now consider admittances of the even and odd states, as well as the absolute value of their difference $Y_{\textnormal{even}} - Y_{\textnormal{odd}}$, which directly translates to the visibility of the readout of the two states \cite{Derakhshan_2020,vonhorstig2024electrical}. 
These quantities are shown in Fig.~\ref{fig:fig4_Majorana} as a function of the QD detuning for the example parameters $\Delta_{\textnormal{even}} = 2.5$ GHz and $\Delta_{\textnormal{odd}} = 0.5$ GHz. 
From Fig.~\ref{fig:fig4_Majorana}e it is clear that in the adiabatic limit, both states have a non-zero admittance, thus creating additional challenges for readout when compared to other qubit systems \cite{Derakhshan_2020}. This, however, can be obviated by measuring at finite frequency. In particular, by choosing a readout frequency $\Delta_{\textnormal{odd}} < \omega < \Delta_{\textnormal{even}}$, one can use the Rabi-induced peak splitting to suppress the admittance of the less-coupled mode (odd in this case) at $\varepsilon = 0$, while the more-coupled mode remains below resonance and still shows a zero-centered peak, as shown in Fig.~\ref{fig:fig4_Majorana}d.
The same panel, furthermore, shows how, while the odd state is indeed dark for $\varepsilon = 0$, the Rabi wings give a strong response for $\varepsilon_{\textnormal{odd}} = \sqrt{\omega - \Delta_{\textnormal{odd}}}$ where the odd subspace is resonant with the excitation. Thus, in this setting, one can selectively choose \textit{which} state is bright by changing the readout voltages. 
This capability has been discussed in the context of other qubit architectures, and has been proposed to have several advantages with respect to more conventional readout techniques based on negative measurements, such as the ability to check for leakage outside the computational basis \cite{vonhorstig2024electrical,lundberg2021nonreciprocal}.
Our formalism thus predicts the possibility to selectively (and dynamically) choose the bright state of a readout measurement by exploiting the diabatic properties of the system's admittance, overcoming the limitation of this platform in the adiabatic regime.

\section*{Conclusion and Outlook}

We have presented a novel Lindblad perturbation theory for quantum modelling of the electrical response of a generic QD device, highlighting the link between the susceptibility of a quantum system and its small-signal electrical admittance, from which stems a universal small-signal equivalent circuit. This condenses the perturbed quantum dynamics into linear circuit components, making use uniquely of frequency-independent variable components, simplifying the implementation of the model and providing key insights into the complex response of the Lindbladian. 
Our model shows how semiclassical and input-output approaches describe two facets of the decohering quantum dynamics, overlapping in the description of the adiabatic and perfectly coherent limit: the quantum capacitance.
Moreover, we introduce a novel contribution to the response, named the \textit{Hermes} admittance, which casts a new and complementary light on the concept of quantum capacitance, which can additionally be seen as the response of the quantum system when decoherence is so fast that all coherent processes are suppressed (i.e., the semiclassical limit).

Our novel formalism provides an intuitive model for the small-signal electrical response of any quantum system, retaining the complexity of Lindblad circuit QED while readily implementable within circuit simulators, showcasing it on two example systems.
We have described the electrical response at high frequency of a DQD charge qubit, demonstrating the ability of the model to continuously transitions from the Sisyphus-broadened adiabatic capacitance to the Rabi-induced peak splitting predicted by circuit QED. The single, zero-centered peak of the quantum capacitance is then recovered when considering fast dephasing thanks to the inclusion of the Hermes term.
The phenomenon of resonant peak splitting has then been exploited to showcase the response of a Majorana qubit composed of two Majorana zero modes coupled to an ancillary QD. The admittance of the QD can be used to read the state of the qubit, and we showcase how the diabatic response of the system can be leveraged to selectively make either state dark by varying the readout detuning. We stress that such a scheme exploits the high-frequency capabilities of the model to get around limitations that cannot be overcome in the adiabatic regime, thus highlighting how accurate modelling of the electrical response of quantum systems is key to unlocking the full potential of upcoming quantum technologies.

\section*{Acknowledgments}

The authors acknowledge G. Burkard and A. P\'alyi for useful discussions.
L.P. acknowledges the UK Engineering and Physical Sciences Research Council (EPSRC) via the Cambridge NanoDTC (EP/L015978/1) and the Winton Programme for the Physics of Sustainability. M.F.G.Z. acknowledges a UKRI Future Leaders Fellowship [MR/V023284/1].
MB acknowledges funding from the Emmy Noether Programme of the German Research Foundation (DFG) under grant no. BE 7683/1-1.

\appendix


\section{Gate Current in Multi-Gate QD Systems}
\label{app:MultiGate}

In this section, we discuss how to generalize the formalism discussed in the main text to a multi-gate system, with particular emphasis on QD arrays.
If we consider the $i$-th gate to be coupled to the $j$-th QD with a capacitance $C_{i,j}$, at equilibrium, the screening charge will read
\begin{equation}
  \frac{Q_{i,j}}{C_{i,j}} = \frac{e P_j}{C_{\Sigma_j} },
\end{equation}
\noindent
where $P_j(t)$ is the probability of the $j$-th QD being occupied, and $C_{\Sigma_j}$ is its self-capacitance. 
It is therefore natural to write \cite{Mizuta2017,vigneau2022probing,Esterli_Otxoa_Gonzalez-Zalba_2019}
\begin{equation}
  I_{{\rm G}_{i,j}}(t) = \alpha_{ij} e \frac{d}{dt} P_j(t),
  \label{eq:I_G}
\end{equation}
\noindent 
as the gate current, which is caused on the $i$-th gate of the array by a charge redistribution event on the $j$-th QD. In \eqr{eq:I_G} we have defined the lever arm  $\alpha_{ij} = C_{i,j} /C_{\Sigma_j}$, which quantifies the gate-QD coupling \cite{vigneau2022probing,Buttiker_1993,Cottet_Mora_Kontos_2011}.

The Hamiltonian for the system reads
\begin{equation}
 H = H_0 + e\delta V_i\Pi_{i} \cos{\omega t},
 \label{eq:H_coupl}
\end{equation} 
\noindent 
where $H_0$ is the unperturbed Hamiltonian of the QD system, and $\Pi_{i}$ is the operator that couples the QD array to the perturbation of the $i$-th gate. 
This, as one would expect, reads 
\begin{equation}
  \Pi_{i} = \sum_j \alpha_{ij} \ket{j}\bra{j},
\end{equation}
\noindent
showing that this is a projector on the charge-localized basis, weighted by the respective capacitive couplings.
This generalizes the formalism of the main text, where the single lever arm was just a multiplicative scalar, and allows us to define the detuning oscillation caused by the $i$-th gate amplitude simply as $\delta \varepsilon_{ij} = \alpha_{ij} e \delta V_i$.

To consider the gate current from a quantum-mechanical perspective, we note that it is the \textit{same} lever arm that both defines the detuning oscillation \textit{and} the gate current collection. Therefore, it is natural to rewrite \eqr{eq:I_G} as 
\begin{equation}
  I_{{\rm G}_{i}}(t) = e \frac{d}{dt} \braket{\Pi_i}(t),
  \label{eq:I_G_q}
\end{equation}
\noindent
where $\braket{\Pi_i}(t)$ indicates the (time-dependent) expectation value of the operator $\Pi_i$. 
Intuitively, this must be the case because in the mesoscopic picture the energy of the electrons in the QDs is dictated by the gate voltage times the \textit{effective} charge $\alpha_{ij}e$ seen by the gate, and it is precisely the movement of this charge (and its back-action on the classical circuit) that we are aiming to describe. Therefore, we can expect the same operator that models the coupling to be the one that collects its back-action.

To link $\Pi_i$ to the Hamiltonian, we consider that electrons are trapped in the QDs by the potential defined by the gates and the resulting Coulomb blockade. Therefore, up to a constant which can be defined as zero, the energy of a fully-localized electron \textit{solely} depends on the electrostatic potential generated by the gates (and the charge they accumulated). If we now perturb this field with an oscillatory voltage (small enough to neither significantly change the shape of the confining potential nor break the blockade), the QD energy will change \textit{linearly} with it. 
Thus, this leads to two very important conclusions. 
Firstly, we can write 
\begin{equation}
  e\Pi_i = \frac{d}{dV_i} H_0.
  \label{eq:Pi_def}
\end{equation}
Secondly, from the linear dependence on voltages (i.e., in the constant interaction approximation), we have 
\begin{equation}
  \frac{d^2}{dV_i^2} H_0 = 0
  \label{eq:linear_coupl}
\end{equation}
\noindent
for every gate. 

In the main text, for clarity of exposition, we consider a single (differential) lever arm $\alpha$ to convert gate voltage to detuning. This is generally not the case for arrays of more than 2 QDs. However, all the results are trivially generalizable to an arbitrary lever-arm matrix with the aforementioned definitions.

\section{Equilibrium Populations}
\label{P_eq}

In this work, we have seen how the calculation of the susceptibility requires knowledge of the equilibrium probability of occupation. Therefore, for the benefit of the reader, we include this short Appendix on the means of calculations of the equilibrium populations of a quantum system.

If the dynamics of the system is already described by an LME via a superoperator $\mathcal{L}_0$, the steady-state density matrix will simply read 
\begin{equation}
  \mathcal{L}_0 \rho_{\rm ss} = 0,
\end{equation}
\noindent
which, for the non-driven system, is simply the thermal equilibrium.
Therefore, finding $\rho_{\rm ss}$ simply turns into an eigenvalue problem of finding the kernel of $\mathcal{L}_0$. This can be easily done, in the general case, in Fock-Liouville space 
\cite{manzano_short_2020,Am-Shallem_Levy_Schaefer_Kosloff_2015}, where it is possible to turn the superoperator eigendecomposition into a matrix eigenvalue problem\cite{Albert_Bradlyn_Fraas_Jiang_2016}.

In some cases, however, the coupling with the environment will be small enough to be negligible in the dynamics within an rf cycle, but nonetheless large enough to dictate the equilibrium steady state in the long-time regime. In this case, as we have already discussed, the density matrix will become diagonal, and one only needs to find the occupation probability in thermal equilibrium. 
The simplest way to do so is to consider the system coupled to a bosonic bath (i.e., phonons in the crystal), which is responsible for $T_1$-type processes.
The Hamiltonian in this case reads
\begin{equation}
  H_{\rm S-B} = \lambda \sum_{E_m > E_n} \tau^x_{mn} b +\tau^x_{nm} b^\dagger
\end{equation}
\noindent
where $b$ is the destruction operator on the bath. 
In this case, the LME reads 
\begin{equation}
  \mathcal{L} \rho = -i \left[H_0, \rho\right] + \mathcal{L}_{T_1} ,
\end{equation}
\noindent with 
\begin{equation}
  \begin{aligned}
    \mathcal{L}_{T_1} =  &\Gamma_0 \sum_{E_m > E_n}N(E_m - E_n) \mathcal{D}\left(\tau^x_{nm} \right) \rho +\\
  + &\Gamma_0 \sum_{E_m > E_n} \big(N(E_m - E_n)+1\big) \mathcal{D}\left(\tau^x_{mn} \right) \rho ,
\end{aligned}
\label{eq:L_T1}
\end{equation}
\noindent
where we assume for the summation that $E_m > E_n$ if $m>n$, and there is no forbidden transition.
In the rotating wave approximation, we simply take \cite{Mizuta2017,kohler_dispersive_2018}
\begin{equation}
  N(E) = \int_{-\infty}^{\infty} \braket{b^\dagger b} (\epsilon) \delta(E-\epsilon) d \epsilon
\end{equation}
\noindent
as the number of bosons resonant with the transition. For simplicity, we take $\Gamma_0$ to be independent of energy and the same for all transitions.
Typically, one could consider a Debye model \cite{kohler_dispersive_2018}
\begin{equation}
  N(E) = \frac{E}{\mathcal{E}_0} \frac{1}{e^{E/k_{\rm B} T} -1},
  \label{eq:bath_number}
\end{equation}
\noindent
where $\frac{1}{e^{E/k_{\rm B} T} -1}$ is the Bose-Einstein distribution and $E/\mathcal{E}_0$ represents the bath spectral density.
On the diagonal of the density matrix, these jump operators take the form of a Pauli-like master equation, reading
\begin{equation}
  \begin{aligned}
  \frac{d}{dt} p_n = \Gamma_0 \bigg(\sum_{m>n} -N_{mn} p_n + (N_{mn}+1) p_m +\\
  + \sum_{m<n} -(N_{nm}+1) p_n  + N_{nm} p_m \bigg)
\end{aligned}
\end{equation}
\noindent
where, for ease of notation, we have defined $N_{mn} = N(E_m - E_n)$. Therefore, the equilibrium probabilities are found as the eigenvalues of the rate matrix 
\begin{equation}
  R_{mn} = \Gamma_0 \begin{cases}
    N_{mn}+1 \hfill &\textnormal{if} ~~E_m > E_n\\
    -\sum_{m} N_{mn} - \sum_{m<n} 1 \hfill &\textnormal{if} ~~ m=n \\
    N_{mn} \hfill &\textnormal{if} ~~E_m < E_n
  \end{cases}
\end{equation}
\noindent 
from which it is clear that the steady-state does not depend on the choice of $\Gamma_0$. Moreover, in the limits of very low ($N_{mn}\ll1$) or very high ($N_{mn}\gg1$) temperature, the steady state does not depend on $\mathcal{E}_0$ \cite{kohler_dispersive_2018,kohler_driven_2005}. 

\section{Numerical Calculation of the Gate Current}
\label{C_param_der}

In this section, we will show how to derive the admittance from Eqs.(\ref{eq:I_G_q_main}-\ref{eq:pi_dec_prov}) for an arbitrary perturbation of a markovian time evolution. In particular, we will consider the most general case of a Lindbladian of the type 
\begin{equation}
  \mathcal{L}(t) = \mathcal{L}_0 + \delta \mathcal{L}(t) ,
\end{equation}
\noindent
where we consider $\delta \mathcal{L}$ to be a small perturbation to $\mathcal{L}_0$. We will then be able to write the time evolution as
\begin{equation}
  \rho(t) = \rho_0 + \delta \rho(t) = \sum_m (p_{m,0} + \delta p_m) \left|\phi_m \rangle \langle \phi_m \right|,
\end{equation}
\noindent
where we assume that $\rho(t)$ mostly remains diagonal and 
\begin{equation}
  \mathcal{L}_0 \rho_0 = 0
\end{equation}
\noindent
for the stationary state.

We must now compute $I_{\rm G}(t)$. We can write, however
\begin{equation}
  I_{\rm G}(t) = \alpha e \frac{d}{dt}\braket{\Pi}(t) =\alpha e \frac{d}{dt}\tr{\Pi \rho(t)}.
\end{equation}
\noindent 
By definition of the Lindbladian, this becomes
\begin{equation}
  I_{\rm G}(t) =\alpha e \tr{\Pi \mathcal{L}(t) \rho(t)} .
\end{equation} 
To first order in the perturbation, this becomes 
 \begin{equation}
  I_{\rm G}(t) =\alpha e \left(\tr{\Pi \delta \mathcal{L}(t) \rho_0} + \tr{\Pi \mathcal{L}_0 \delta \rho(t)} \right),
  \label{eq:I_G_lind_pert}
 \end{equation}
\noindent
echoing the fact that $\delta \rho(t)$ is, to first order in the perturbation, the solution of the equation
\begin{equation}
  \frac{d}{dt} \delta \rho(t) = \delta \mathcal{L}(t) \rho_0 + \mathcal{L}_0 \delta \rho(t).
  \label{eq:small_rho_app}
\end{equation}

It is interesting to point out how \eqr{eq:small_rho_app} is the direct Lindblad generalization to the expression of tunnelling capacitance presented in Ref.~\cite{lundberg2021nonreciprocal}. However, the necessity to solve a complex time-dependent differential equation to obtain $\delta \rho(t)$ makes this formulation only viable for numerical treatments. 

\section{Derivation of the Susceptibility}

In this section, we shall sketch for completeness and pedagogical value the necessary mathematical steps to derive the susceptibility formulas in the main text from the general expression in \eqr{eq:chi_def_L} (and \eqr{eq:chi_def}).

\subsection{Hamiltonian Susceptibility}
\label{Chi_calc}

The first term we tackle is the case of the isolated system.
As mentioned in the main text, \eqr{eq:chi_H} is equivalent to Ref.\cite{kohler_dispersive_2018}. However, we report it here for completeness and to illustrate the physical origin of the imaginary response in \eqr{eq:chi_L}, as well as it naturally extending to describe $\chi_{\rm H}$ in \eqr{eq:chi_L}.

To begin with, we can define
\begin{equation}
  U(t, \tau) = \exp{ \left( -i H_0(t-\tau)\right)} 
\end{equation}
\noindent 
through which the operators evolve in the interaction picture as
\begin{equation}
  \Pi(t, \tau) = U^\dagger(t, \tau) \Pi U(t, \tau). 
\end{equation}
Therefore, making the commutator explicit, 
\begin{equation}
  \begin{aligned}
  \chi(t, \tau) &= -i \big(\tr{U^\dagger(t, \tau) \Pi U(t, \tau) \Pi \rho_{\rm ss}}-\\
  -&\tr{\Pi U^\dagger(t, \tau) \Pi U(t, \tau) \rho_{\rm ss}}\big) \Theta(t-\tau).
\end{aligned}
\label{eq:chi_explicit}
\end{equation}

We can now take the case when $\rho_{\rm ss}$ is purely diagonal. This may not be the case for a perfectly Hamiltonian dynamics. However, \textit{any} interaction with the environment will lead to exponentially decaying off-diagonal terms \cite{kohler_floquet-markovian_1997}. Therefore, we can consider this as the limit of $\Gamma^{mn}_{T_2} \rightarrow 0$.
For a practical way of computing such probabilities, a method is presented in App.~\ref{P_eq}.
In this case, we can use the fact that 
\begin{equation}
  \rho_{\rm ss} = \sum_m p _m \left| \phi_m \rangle \langle  \phi_m \right|
\end{equation}
\noindent
and the cyclic property of the trace to write the first term in \eqr{eq:chi_explicit} as
\begin{equation}
  \begin{aligned}
  \tr{U^\dagger(t, \tau) \Pi U(t, \tau) \Pi \rho_{\rm ss}}= \\
  = \sum_m p_m \braket{\phi_m \left| U^\dagger(t, \tau) \Pi U(t, \tau) \Pi\right|\phi_m}.
\end{aligned}
\label{eq:chi_trace_middle}
\end{equation}
Now, we can use the completeness relation
\begin{equation}
  \mathcal{I} = \sum_n  \left| \phi_n \rangle \langle  \phi_n \right|,
\end{equation}
\noindent 
where $\mathcal{I}$ is the identity, to obtain
\begin{equation}
  \sum_{m,n} p_m \braket{\phi_m \left| U^\dagger(t, \tau) \Pi U(t, \tau) \right|\phi_n} \braket{\phi_n \left| \Pi\right|\phi_m}.
\end{equation}
Performing similar manipulation for the other term in \eqr{eq:chi_explicit} and making the time evolution of bra and ket explicit, we find
\begin{equation}
  \begin{aligned}
  \chi(t, \tau) = -i \sum_{m,n} (p_m - p_n) e^{i (E_m - E_n)(t-\tau)} \cdot \\
  \cdot |\braket{\phi_n \left| \Pi\right|\phi_m}|^2 \Theta(t-\tau),
  \label{eq:chi_tt}
\end{aligned}
\end{equation}
\noindent
whose Fourier transform is \eqr{eq:chi_H}. 

This derivation allows to immediately consider the expression for $\chi_{\rm H}(t, \tau)$ in the case of the Lindblad dynamics. 
Using the cyclic property of the trace, in fact, we can write the equivalent of \eqr{eq:chi_tt} as 
\begin{equation}
  \begin{aligned}
  \chi_{\rm H}(t, \tau) = -i \sum_{m,n} (p_m - p_n) e^{i (E_m - E_n)(t-\tau)} \cdot \\
  \hfill \cdot  e^{-\Gamma_{T_2}^{mn} (t-\tau)}  \hfill |\braket{\phi_n \left| \Pi\right|\phi_m}|^2 \Theta(t-\tau),
\end{aligned}
\end{equation}
\noindent
where we have defined $\Gamma_{T_2}^{mn}=\frac{1}{2}\Gamma_{T_1}^{mn} + \Gamma_{\phi}^{mn}$ as in the main text, and used the fact that $[\Pi, \rho_{\rm ss}]$ is real and antihermitian, and thus non-zero only off the diagonal, as well as
\begin{equation}
  \mathcal{L}_0 \big|\phi_m \rangle \langle \phi_n \big| = \left( i (E_m - E_n) - \Gamma^{mn}_{T_2} \right) \big|\phi_n \rangle \langle \phi_m \big|.
  \label{eq:decay}
\end{equation}
Taking the Fourier transform, we obtain \eqr{eq:chi_L}.

\subsection{Sisyphus Susceptibility}
\label{Sisyphus}

To derive the general form of the Sisyphus admittance, we must write down the variation of $T_1$ with respect to energy. 
Assuming without loss of generality a simple form of \eqr{eq:L_T1}, we can write
\begin{equation}
  \begin{aligned}
  \delta \mathcal{L}_{\Gamma_{T1}} \rho_{\rm ss} = \sum_{E_m > E_n} &\frac{d\Gamma_+^{mn}}{d \varepsilon} \mathcal{D}\left(\tau^x_{mn}\right) \rho_{\rm ss} + \\
   + &\frac{d\Gamma_-^{mn}}{d \varepsilon} \mathcal{D}\left(\tau^x_{nm}\right)\rho_{\rm ss} ,
\end{aligned}
\end{equation}
\noindent
where, for compactness, we write
\begin{equation}
  \frac{d\Gamma_\pm^{mn}}{d \varepsilon} = \frac{d E}{d \varepsilon} \frac{d}{d E} \Gamma_\pm^{mn} \bigg|_{E = E_m - E_n}.
\end{equation}
We can now notice that, for a diagonal density matrix,
\begin{equation}
  \mathcal{D}\left(\tau^x_{nm} \right) \rho_{\rm ss} = \sqrt{2} p_m \tau^z_{mn}.
\end{equation}

\textit{En passant}, we note that the $\delta \mathcal{L}_\Gamma$ due to pure dephasing processes vanishes over a diagonal steady-state $\rho_{\rm ss}$. Therefore, even if the dephasing rate has an energy dependence, it only contributes to $\Gamma^{mn}_{T_2}$ in \eqr{eq:chi_L}, but does not introduce additional terms in the admittance.

Therefore,
\begin{equation}
  \delta \mathcal{L}_{\Gamma_{T1}} \rho_{\rm ss} =  \sqrt{2} \sum_{E_m>E_n}  
  \left(\frac{d\Gamma_+^{mn}}{d \varepsilon} p_m-
  \frac{d\Gamma_-^{mn}}{d \varepsilon} p_n\right) \tau^z_{mn}.
  \label{eq:pet_delta_L_T1}
\end{equation}
If we now define 
\begin{equation}
  \Gamma_{T_1}^{mn} = \Gamma_+^{mn} + \Gamma_-^{mn}
\end{equation}
\noindent
using the definitions in \eqr{eq:Y_sis}, and making use of Eqs. (\ref{eq:pet_delta_L_T1}) and (\ref{eq:chi_def_L}), we easily obtain
\begin{equation}
  \begin{aligned}
  \chi_{T_1}(t, \tau) = \sum_{E_m>E_n} e^{-\Gamma_{T_1}^{mn}(t-\tau)} \sigma_{mn}.
\end{aligned}
\label{eq:chi_sis_g0_const}
\end{equation}
\noindent
whose Fourier transform reads
\begin{equation}
  \begin{aligned}
  \chi_{T_1}(\omega) =\sum_{E_m>E_n}& \big(\delta \Gamma_{mn}^+p_m-\delta \Gamma_{mn}^-p_n\big)  \\
  &\frac{\braket{\phi_n\big|\Pi\big|\phi_n} - \braket{\phi_m\big|\Pi\big|\phi_m}}{\Gamma_{T_1}^{mn} + i \omega}
\end{aligned}
\end{equation}
and 
\begin{equation}
  \delta \Gamma_{mn}^\pm = \frac{dE}{d \varepsilon} \frac{d}{dE} \Gamma^\pm (E) \bigg|_{E = E_m - E_n}.
\end{equation}

\subsection{Hermes Susceptibility}
\label{Hermes}

The first step towards deriving the Hermes admittance is to expand the rotation matrix $W(t)$, which diagonalizes the Hamiltonian instantaneously, to first order. To do so, we can notice that 
\begin{equation}
  W(t) = \sum_m \ket{\phi_m(t)} \bra{\phi_m(t)}.
\end{equation}
Instantaneously, to first order,
\begin{equation}
  \begin{aligned}
  \ket{\phi_m(t)} = \ket{\phi_m} + \sum_{n \neq m} \frac{\braket{\phi_n|\Pi|\phi_m}}{E_m - E_n} \ket{\phi_n} \delta \varepsilon \cos{\omega t} + \mathcal{O}(\delta \varepsilon^2).
\end{aligned}
\end{equation}
Thus, 
\begin{equation}
  W(t) = W^0 + \delta \varepsilon \delta W  \cos{\omega t} + \mathcal{O}(\delta \varepsilon^2),
\end{equation}
\noindent
with 
\begin{equation}
  \delta W = \left(\sum_m \sum_{n \neq m} \frac{\braket{\phi_n|\Pi|\phi_m}}{E_m - E_n} \ket{\phi_n} \bra{\phi_m} - h.c. \right).
\end{equation}
Consequently, the perturbation of the jump operators reads
\begin{equation}
  \begin{aligned}
    \frac{d L_l}{d \varepsilon} &= \delta W^\dagger L_l^{\rm IEB} W^0+ \left(W^0\right)^\dagger L_l^{\rm IEB} \delta W\\
  & =- \left[ \delta W, L^0_l\right].
\end{aligned}
\end{equation}
With some more algebra, it is possible to write 
\begin{equation}
  \begin{aligned}
  \mathcal{D}'(L) = &- \left[\delta W, L^0 \rho \left(L^0\right)^\dagger \right] + \frac{1}{2} \left\{\left[\delta W, \left(L^0\right)^\dagger L^0 \right], \rho\right\}+\\
  & +L^0 \left[\delta W, \rho \right] \left(L^0\right)^\dagger .
  \end{aligned}
\end{equation}
\noindent

We must now consider relaxation and dephasing separately. The Hermes susceptibility of $T_1$ processes can be calculated making use of the algebra presented in App.~\ref{Sisyphus}, with the additional fact that 
\begin{equation}
  \left[\delta W, \ket{\phi_m} \bra{\phi_m}\right] = \sum_{n\neq m } \frac{\braket{\phi_n|\Pi|\phi_m}}{E_m - E_n} \ket{\phi_n} \bra{\phi_m} + h.c.
  \label{eq:dV_comm}
\end{equation}
With a similar formalism, it is also possible to include pure dephasing ($T_\phi$) processes, which are usually described by a diagonal jump operator $\tau^z_{mn}$. Carrying out the algebra and making use of \eqr{eq:dV_comm}, we find the results in \eqr{eq:d_p_on_rho}.

\begin{widetext}
\begin{equation}
  \begin{aligned}
  &\mathcal{D}'(\tau^x_{nm} ) \rho_{\rm ss} = - \sum_{n\neq m } \frac{3p_m+p_n}{2} \left(\frac{\braket{\phi_n|\Pi|\phi_m}}{E_n - E_m} \ket{\phi_n} \bra{\phi_m} - \frac{\braket{\phi_m|\Pi|\phi_n}}{E_m - E_n} \ket{\phi_m} \bra{\phi_n}\right) = - \sum_{n\neq m } \frac{3p_m+p_n}{2} \Lambda_{mn} \\
  &\mathcal{D}'\left(\tau^z_{nm}\right) \rho_{\rm ss} = \sum_{n\neq m } (p_m-p_n) \left(\frac{\braket{\phi_n|\Pi|\phi_m}}{E_n - E_m} \ket{\phi_n} \bra{\phi_m} - \frac{\braket{\phi_m|\Pi|\phi_n}}{E_m - E_n} \ket{\phi_m} \bra{\phi_n}\right) = \sum_{n\neq m } (p_m-p_n) \Lambda_{mn} .
  \end{aligned}
  \label{eq:d_p_on_rho}
\end{equation}
\end{widetext}

Notably, one can express both superoperators in terms of $\Lambda_{mn}$, and they only differ in their dependence on the probabilities. The formulae greatly simplify if we consider the combined action of relaxation and dephasing. In the notation of App.~\ref{Sisyphus} and taking, without loss of generality, $E_m > E_n$, the perturbation to the jump operators reads
\begin{equation}
  \begin{aligned}
  \delta \mathcal{L}_{\rm L}^{mn} \rho_{\rm ss} &= \Gamma_\phi^{mn} \mathcal{D}'\left(\tau^z_{nm}\right) \rho_{\rm ss} +\\
  & + \Gamma_+^{mn} \mathcal{D}'\left(\tau^x_{mn}\right) \rho_{\rm ss}
+ \Gamma_-^{mn} \mathcal{D}'\left(\tau^x_{nm}\right) \rho_{\rm ss},
  \end{aligned}
\end{equation}
\noindent
where we have introduced the pure dephasing rate $\Gamma_\phi$.

Before proceeding, we note that in such a Lindbladian the equilibrium probabilities and the relaxation rates are linked by the principle of detailed balance as
\begin{equation}
  \frac{p_m}{p_n} = \frac{\Gamma_+^{mn}}{\Gamma_-^{mn}}.
  \label{eq:detaileb_balance}
\end{equation}

Combining Eqs.~(\ref{eq:d_p_on_rho}) and (\ref{eq:detaileb_balance}) we find, after some algebra, 
\begin{equation}
  \delta \mathcal{L}_{\rm L}^{mn} \rho_{\rm ss} = \Gamma_{T_2}^{mn} (p_m - p_n) \Lambda_{mn},
  \label{eq:Herm_lind_partial}
\end{equation}
\noindent
where we have defined 
\begin{equation}
  \Gamma_{T_2}^{mn} = \Gamma_{\phi}^{mn} + \frac{1}{2} \Gamma_{T_1}^{mn}.
\end{equation}

Interestingly, there is a factor of 2 between the contributions of relaxation and dephasing, reflecting the fact that the decay of the coherences reads $T_2^{-1} = \left(T_\phi\right)^{-1} + \left(2T_1\right)^{-1}$.

From \eqr{eq:Herm_lind_partial} it follows directly that we can write the Hermes susceptibility as
\begin{equation}
  \begin{aligned}
  \chi_{\rm L}(t, \tau) = i\sum_{m \neq n} \Gamma_{T_2}^{mn} \frac{p_m - p_n}{E_m - E_n} e^{i (E_m - E_n)(t-\tau)} \cdot \\
  \hfill \cdot  e^{-\Gamma_{T_2}^{mn} (t-\tau)}  \hfill |\braket{\phi_n \left| \Pi\right|\phi_m}|^2 \Theta(t-\tau)
\end{aligned}
\end{equation}
\noindent
the Fourier transform of which is \eqr{eq:chi_HE} in the main text.


%

\end{document}